\documentclass[twocolumn,aps,prd,preprintnumbers,superscriptaddress,nofootinbib,amsmath,amssymb,floats,floatfix,showkeys,notitlepage,showpacs]{revtex4-1}

\usepackage{orcidlink}
\usepackage{comment}
\usepackage{lipsum}
\usepackage{graphicx}
\usepackage{subfigure}
\usepackage{palatino}
\usepackage{sans}
\usepackage{hyperref}
\hypersetup{colorlinks=true,linkcolor=blue,urlcolor=blue,citecolor=blue}
\usepackage[toc,page]{appendix}
\usepackage[normalem]{ulem}
\usepackage{adjustbox}
\usepackage{latexsym}
\usepackage{amsmath}
\usepackage{amssymb}
\usepackage{amsfonts}
\usepackage{dcolumn}
\usepackage{bm}
\usepackage{tikz}
\usepackage{bigints}
\usepackage{array,tabularx,multirow,booktabs}
\usepackage[tracking=true]{microtype}
\SetTracking{}{500}
\SetTracking{encoding={*}, shape=sc}{40}
\UseRawInputEncoding 
\allowdisplaybreaks

\begin{document}
	\newcommand \nn{\nonumber}
	\newcommand \fc{\frac}
	\newcommand \lt{\left}
	\newcommand \rt{\right}
	\newcommand \pd{\partial}
	\newcommand \e{\text{e}}
	\newcommand \hmn{h_{\mu\nu}}
	\newcommand{\PR}[1]{\ensuremath{\left[#1\right]}} 
	\newcommand{\PC}[1]{\ensuremath{\left(#1\right)}} 
	\newcommand{\PX}[1]{\ensuremath{\left\lbrace#1\right\rbrace}} 
	\newcommand{\BR}[1]{\ensuremath{\left\langle#1\right\vert}} 
	\newcommand{\KT}[1]{\ensuremath{\left\vert#1\right\rangle}} 
	\newcommand{\MD}[1]{\ensuremath{\left\vert#1\right\vert}} 

	
\title{Weak field deflection angle and analytical parameter estimation of the Lorentz-violating Bumblebee parameter through the black hole shadow using EHT data}

\author{Gaetano Lambiase \orcidlink{0000-0001-7574-2330}}
\email{lambiase@sa.infn.it}
\affiliation{Dipartimento di Fisica ``E.R Caianiello'', Università degli Studi di Salerno, Via Giovanni Paolo II, 132 - 84084 Fisciano (SA), Italy.}
\affiliation{Istituto Nazionale di Fisica Nucleare - Gruppo Collegato di Salerno - Sezione di Napoli, Via Giovanni Paolo II, 132 - 84084 Fisciano (SA), Italy.}

\author{Reggie C. Pantig \orcidlink{0000-0002-3101-8591}}
\email{rcpantig@mapua.edu.ph}
\affiliation{Physics Department, Map\'ua University, 658 Muralla St., Intramuros, Manila 1002, Philippines}

\author{Ali \"Ovg\"un \orcidlink{0000-0002-9889-342X}}
\email{ali.ovgun@emu.edu.tr}
\affiliation{Physics Department, Eastern Mediterranean University, Famagusta, 99628 North
Cyprus via Mersin 10, Turkiye.}

\begin{abstract}

We explored how the Lorentz symmetry breaking parameter $\ell$ affects the Reissner-Nordst\"{o}m BH solution in the context of weak field deflection angle, and the black hole shadow. We aim to derive the general expression for the weak deflection angle using the non-asymptotic version of the Gauss-Bonnet theorem, and we presented a way to simplify the calculations under the assumption that the distance of the source and the receiver are the same. \textcolor{black}{Through the Solar System test, $\ell$ is constrained from around $-10^{-9}$ orders of magnitude to $0$, implying challenging detection of $\ell$ through the deflection of light rays from the Sun.} We also studied the black hole shadow in an analytic way, where we applied the EHT results under the far approximation in obtaining an estimate expression for $\ell$. Using the realistic values of the black hole mass and observer distance for Sgr. A* and M87*, it was shown that $M/r_{\rm o} \neq 1$ is satisfied, implying the relevance and potential promise of the spontaneous Lorentz symmetry breaking parameter's role on the shadow radius uncertainties as measured by the EHT. \textcolor{black}{We find constraints for $\ell$ to be negatively valued, where the upper and lower bounds are $\sim -1.94$ and $\sim -2.04$, respectively.}
\end{abstract}	



\maketitle


\textbf{Introduction:}
Black holes, born from the violent collapse of colossal stars, are foundational components of the cosmos. Governed by the principles of Einstein’s general relativity, these objects exhibit a surprising simplicity given their complex origins. Defined exclusively by their mass, spin, and electric charge—a concept encapsulated by the no-hair theorem—black holes stand as fundamental building blocks of the universe.  General relativity's predictions concerning black holes have profoundly influenced modern physics. Landmark achievements, such as the LIGO/Virgo collaboration's detection of gravitational waves from colliding black holes \cite{LIGOScientific:2016aoc,Poggiani:2024aat} and the Event Horizon Telescope's (EHT) visualization of the supermassive black hole M87*, have inaugurated a new era of black hole exploration \cite{EventHorizonTelescope:2019dse,EventHorizonTelescope:2022wkp}. Investigating the intrinsic characteristics of these enigmatic objects holds the promise of uncovering observable properties that can be subjected to rigorous scientific testing. A small charge of the magnitude notably alters the position of the innermost stable circular orbit for charged particles. Furthermore, the authors introduce a new observational test focusing on the decrease in bremsstrahlung surface brightness, which is more sensitive to smaller unshielded electric charges than the size of the black hole's shadow
\cite{Zajacek:2018ycb}.
\textcolor{black}{The Bumblebee gravity model is well-suited for exploring deviations from General Relativity (GR), which, despite its success, struggles with spacetime singularities, the black hole information paradox, and lacks a complete quantum theory \cite{Kostelecky:1988zi,Kostelecky:1989jw}. GR has been extensively tested in weak-field regimes, but strong-field tests around compact objects are needed. Bumblebee gravity, with its spontaneous Lorentz symmetry breaking, offers exact black hole solutions that differ from GR, impacting observable phenomena like black hole shadows and deflection angles. These features provide a testable framework to probe gravity in strong fields, aligning with experiments such as the Event Horizon Telescope.}

In Bumblebee gravity theory, the Lagrange density preserves Lorentz invariance, but local Lorentz symmetry is spontaneously violated by a potential \(V(B_{\mu} B^{\mu} \mp \bar{b}^{2})\), where \(B^{\mu}\) is a dynamic vector field called the bumblebee field. This leads to spacetime anisotropy and preferred frames, with the underlying geometry assumed to be Riemannian. The model, inspired by a framework introduced by Kostelecky and Samuel in 1989 \cite{Kostelecky:1988zi,Kostelecky:1989jw}, features tensor-induced spontaneous Lorentz symmetry breaking and can be explored in various spacetime geometries, including Riemann-Cartan spacetimes \cite{Kostelecky:2000mm,Kostelecky:2002ca,Bertolami:2003qs}. Significant progress has been made in finding black hole solutions within this theory, including exact solutions for Schwarzschild-like black holes, traversable wormholes, and slowly rotating Kerr-like black holes, etc.  \cite{Santos:2014nxm,Casana:2017jkc,Ding:2019mal,Jha:2020pvk,Filho:2022yrk,Xu:2022frb,Ding:2023niy,AraujoFilho:2024ykw}. \textcolor{black}{While Bumblebee gravity provides distinct predictions for phenomena like black hole shadows, comparisons with alternative beyond-GR theories—such as Einstein-aether models and massive gravity—are needed. Both Bumblebee and Einstein-aether models involve spontaneous Lorentz symmetry breaking, but Einstein-aether includes a dynamic “aether” field interacting with spacetime, offering more flexibility in different regimes \cite{Jacobson:2007veq,Eling:2006ec}. Bumblebee gravity, with its simpler vector field, is mathematically easier but less phenomenologically rich. Massive gravity, on the other hand, modifies GR by giving the graviton a mass, addressing large-scale cosmological issues rather than local Lorentz violations \cite{deRham:2010kj,Hinterbichler:2011tt}. While Bumblebee gravity is more suited for high-energy or small-scale phenomena like black holes, massive gravity deals with large-distance effects. Bumblebee also shares features with the Standard-Model Extension (SME), which systematically introduces Lorentz violations across various interactions, but focuses specifically on the gravitational sector. Each theory offers unique insights, with Bumblebee gravity standing out for its simplicity and focus on local Lorentz symmetry breaking.}

Einstein's general relativity posits that massive bodies bend light, a phenomenon known as gravitational lensing, which is crucial for testing alternative gravity theories. Virbhadra demonstrated that observing relativistic image formations can provide precise upper limits on the compactness of massive, dark entities, independent of their mass and distance \cite{Virbhadra:1999nm,Virbhadra:2002ju,Virbhadra:1998dy,Virbhadra:2007kw}. Gibbons and Werner introduced a novel method for computing weak deflection angles using the Gauss-Bonnet theorem on the optical metric of asymptotically flat black holes \cite{Gibbons:2008rj}. This method has been was extended to accommodate both asymptotically flat and non-asymptotically flat spacetimes by the authors \cite{Ishihara:2016vdc, Li:2020wvn}. Since then, this powerful method has become the interest of many works in the literature \cite{Li:2020wvn,Ishihara:2016vdc} due to its versatility on the different types of spacetime metric.

On the other hand, a distinguishing feature of compact objects such as black holes is their strong gravitational influence on light \cite{Zakharov:2014lqa,Tsukamoto:2017fxq,Cunha:2018acu,Perlick:2021aok,Lima:2021las}. This curvature of spacetime creates an innermost photon orbit, a precarious equilibrium where light particles can neither fall into the black hole nor escape to infinity, resulting in the visible shadow cast against the background of space \cite{Claudel:2000yi}. The Event Horizon Telescope's landmark photos of M87* and Sgr A* provided empirical support for this theoretically expected phenomenon \cite{EventHorizonTelescope:2019dse,EventHorizonTelescope:2022wkp, EventHorizonTelescope:2021dqv}. Despite extensive investigation before and after these observations, the black hole shadow remains a key study topic, offering essential insights into the fundamental spacetime geometry and the influence of surrounding astrophysical environments \cite{Vagnozzi:2019apd,Vagnozzi:2022moj,Chen:2022nbb,Atamurotov:2013sca,Abdikamalov:2019ztb,Abdujabbarov:2016efm,Toshmatov:2021fgm,Abdujabbarov:2016hnw,Atamurotov:2015nra,Rayimbaev:2024fwp,Kumar:2018ple,Alloqulov:2024olb,Sarikulov:2022atq,Li:2022eue}. \textcolor{black}{The optical properties of black holes, such as the photon orbits, shadow size, and deflection angle are significantly influenced by different spacetime parameters across various black hole models. While the literature is too many to mention, studies showed how specific parameters arising from these black hole models leave imprints on certain black hole properties. To show the relevance of these parameters, tight constraints were studied using the Schwarzschild deviation parameter as reported by the EHT. Constraining these parameters is so important for distinguishing between different gravity models and testing the validity of alternative theories, as they directly affect observable quantities like black hole shadows. By accurately measuring these optical properties through astronomical observations, we can place limits on these parameters, enhancing our understanding of black hole physics and the fundamental structure of spacetime.}


Our aim in this work is to investigate the effect of the Lorentz-violating parameter $\ell$ on the weak field deflection angle in its most general form where finite distance of the source and receiver are involved and with the deflection of massive particles. The calculations are subtle, but we introduce here a more simplified way of deriving the expression. Next is also investigating how $\ell$ affects the shadow of the black hole, where we also derive analytic formulas for it. 

\textbf{Brief Review of the Charged SSBH in Bumblebee theory:} 
The bumblebee model extends general relativity by incorporating the bumblebee field $B_{\mu}$, a vector field that interacts nontrivially with gravity and acquires a non-zero vacuum expectation value through a specific potential. This causes a spontaneous violation of Lorentz symmetry within the gravitational sector, as described by Kostelecky \cite{Kostelecky:2003fs}
$S=\int d^{4}x\sqrt{-g} \bigg [  \frac{1}{2\kappa}\left(R\right)+\frac{\xi}{2\kappa}B^{\mu}B^{\nu}R_{\mu\nu}-\frac{1}{4} B_{\mu\nu}B^{\mu\nu}-V(B^{\mu}B_{\mu}\pm \bar{b}^{2}) \bigg ] +\int d^{4}x\sqrt{-g}\mathcal{L}_{M},.$
The gravitational coupling constant is $\kappa=\frac{8\pi G}{c^4}$, while $\xi$ represents the non-minimal coupling constant between gravity and the bumblebee field. Similar to the electromagnetic field, the bumblebee field strength is defined as $B_{\mu\nu}=\partial_{\mu}B_\nu-\partial_{\nu}B_{\mu}.$

We consider the matter field as an electromagnetic field that is non-minimally coupled with the bumblebee vector field \cite{Lehum:2024ovo}, and the Lagrangian density is given by the equation:
\begin{eqnarray}
	\mathcal{L}_{M}&=&\frac{1}{2\kappa}(F^{\mu\nu}F_{\mu\nu}+{\gamma}B^{\mu}B_{\mu}F^{\alpha\beta}F_{\alpha\beta}).
	\label{L}
\end{eqnarray} 
Note that electromagnetic  (EM) tensor related to the electromagnetic field (EF) is $	F_{\mu\nu}=\partial_{\mu}A_\nu-\partial_{\nu}A_{\mu}$and $ A =(\phi(r),0,0,0) $
with coupling coefficient ${\gamma}$. Note that the potential $V$ induces a non-zero vacuum expectation value $B_\mu$ through a functional form dependent on $V(B^{\mu}B_{\mu}\pm \bar{b}^{2})$ and necessitates the condition $\bar{b}^{\mu}\bar{b}_{\mu}=\mp \bar{b}^{2}=\mathrm{const}$ where $\left\langle B_{\mu}\right\rangle = \bar{b}_{\mu}$. Also one can define $X=B^{\mu}B_{\mu}\pm \bar{b}^{2}$ and
$V'=\frac{\partial V}{\partial X}$.

By varying the action and solving the resulting field equations, the metric for static and spherically symmetric black holes can be expressed as follows:
\begin{equation} \label{ds^2}
	{d}{s}^{2}=-A(r) {dt}^{2}+B(r
	){dr}^{2}+C(r) \left({~d}\theta^{2}+ \sin ^{2} \theta {d} \phi^{2}\right),
\end{equation}
and we use the bumblebee field as given by~\cite{Bertolami:2005bh,Casana:2017jkc}
$B_{\mu}=\bar{b}_{\mu}$
and $\bar{b}_{\mu}=\left(0,\bar{b}_{r}(r),0,0\right)$ which gives $\bar{b}_{\mu}\bar{b}^{\mu}=\bar{b}^{2}={\rm const}$, and
$\bar{b}_r(r)=\bar{b}\sqrt{B(r)}.$ Note that $\ell=\xi \bar{b}^2$ is a Lorentz-violating
parameter. Using the conditions $V=0$ and $V'=0$ given in \cite{Casana:2017jkc}, potential is chosen as \cite{Liu:2024axg}
$V(X)=\frac{\gamma}{2}X^2$
with a constant $\gamma$.

The black hole solutions are found as follows: \cite{Liu:2024axg}
\begin{align} \label{coeff}
	A(r)&=1-\frac{2 M}{r}+\frac{2(1+\ell) Q_{0}^2}{(2+\ell) r^2} = 1-\frac{2 M}{r} + \frac{k Q_0^2}{r^2}, \nn \\
	B(r)&=\frac{1+\ell}{A(r)} = \frac{\lambda^2}{A(r)}, \nn \\
	C(r) &= r^2, \qquad \phi(r)=\frac{Q_{0}}{r}.
\end{align}

Note that bumblebee parameter can be written as $k=\frac{2(1+\ell)}{(2+\ell)}$ and $\lambda=\sqrt{1+\ell}$.
This metric, which reduces to the Reissner-Nordstr\"{o}m solution when the parameter $\ell$ is zero, exhibits asymptotic non-Minkowskian behavior as the metric functions \(A(r)\) and \(B(r)\) approach 1 and \(1 + \ell\) at infinity, and it also features two horizons like the Reissner-Nordstr\"{o}m black hole: $r_{\pm}=M \pm \sqrt{M^2-kQ_0^2}.$
In this context, it is evident that the mass and charge parameters of the black hole must satisfy the following condition:$\frac{ Q_0^2}{M^2}\leq k.$
\\ \\
\textbf{Weak Deflection Angle using Gauss-bonnet theorem}: 
Li et al. (2020) applied the Gauss-Bonnet theorem (GBT) to black holes without asymptotic flatness \cite{Li:2020wvn}. While our metric is asymptotically flat, we can adapt their method. \textcolor{black}{ Using the photonsphere radius  $r_{\rm ph}$ as a boundary for integration (by solving the equation
$   A(r_{\rm ph})'r_{\rm ph}^2 - 2A(r_{\rm ph})r_{\rm ph} = 0$ for $r_{\rm ph}$} \cite{Vagnozzi:2022moj}), we calculate the weak deflection angle $
    \Theta = \iint_{_{r_{\rm ph}}^{R }\square _{r_{\rm ph}}^{S}}KdS + \phi_{\text{RS}}.$
\textcolor{black}{Note that $S$ and $R$ in the quadrilateral denote the radial positions of the source and receiver, respectively. Moreover, part of the integration domain is the radius of the photonsphere, denoted by $r_{\rm ph}$, instead of $\infty$. With the photonsphere, it is explained in Ref. \cite{Li:2020wvn} how it resolves the apparent divergence of the Gaussian curvature $K$ (a geometric property that measures the curvature of a surface at a given point) if $\infty$ is used.} The Gaussian curvature is calculated using information about the surface's intrinsic properties, specifically the affine connections (which describe how vectors change along curves on the surface) and the determinant of the metric tensor (which measures distances and angles on the surface). A positive $K$ indicates a spherical-like curvature, a negative $K$ indicates a saddle-like curvature, and a zero $K$ implies a flat surface.

The infinitesimal surface element $dS$ is defined as the square root of the metric determinant multiplied by $dr$ and $d\phi$ as $ dS = \sqrt{g}drd\phi.$ The coordinate angle between source and receiver is $\phi_{\rm RS} = \phi_{\rm R}-\phi_{\rm S}$, where $\phi_{\rm R} = \pi - \phi_{\rm S}$. The Jacobi metric determinant is denoted by $g$. Using the metric \eqref{ds^2}, we define the Jacobi metric with the energy of the massive particle  
$E = \frac{\mu}{\sqrt{1-v^2}},$ and $
    dl^2=g_{ij}dx^{i}dx^{j}
    =(E^2-\mu^2A(r))\left(\frac{B(r)}{A(r)}dr^2+\frac{C(r)}{A(r)}d\Omega^2\right).$

The particle's energy $E$ depends on its velocity $v$. By spherical symmetry, we can study the equatorial plane without loss of generality. The Jacobi metric determinant can be calculated as 
$ g=\frac{B(r)C(r)}{A(r)^2}(E^2-\mu^2 A(r))^2.$
Following Li et al.'s approach  \cite{Li:2020wvn}, the weak deflection angle is computed by integrating over the quadrilateral bounded by the photon sphere, the source, and the receiver positions
$\Theta = \int^{\phi_{\rm R}}_{\phi_{\rm S}} \int_{r_{\rm ph}}^{r(\phi)} K\sqrt{g} \, dr \, d\phi + \phi_{\rm RS}.$

Next, we derive the orbit equation. For massive particles, this follows from the geodesic equation, leading to  $g_{\mu \nu}dx^{\mu}dx^{\nu} = -1$ and 
Here, we substituted $r = 1/u$ and used the particle's angular momentum $J = E v b$, where $b$ is the impact parameter. Using the metric components, we obtain $   F(u) =   \left(\frac{du}{d\phi}\right)^2 =  \frac{1}{(1+\ell)} \left \{ {\frac {{E}^{2}}{{J}^{2}}}- \left( k{Q_0}^{2}{u}^{2}-2\,Mu+1 \right) 
 \left( {J}^{-2}+{u}^{2} \right)
 \right \}.$ 

To find an approximate solution to the equation, we employ an iterative method. This is necessary because a closed-form solution is intractable. Through iterative solving, we determine
$ u\left(\frac{\phi}{\lambda} \right) = \frac{1}{b}\sin\left(\frac{\phi}{\lambda}\right)+\frac{1+v^2\cos^2\left(\frac{\phi}{\lambda}\right)}{b^2v^2}M - \frac{kQ_0^2}{2v^2 b^3}.$

The Gaussian curvature $K$ is expressed in terms of affine connections and the determinant g as follows:
$ K=-\frac{1}{\sqrt{g}}\left[\frac{\partial}{\partial r}\left(\frac{\sqrt{g}}{g_{rr}}\Gamma_{r\phi}^{\phi}\right)\right]$ 
where $\Gamma_{rr}^{\phi} = 0$. Using the analytically derived expression for the photonsphere radius, $r_{\rm ph}$, $
    \left[\int K\sqrt{g}dr\right]\bigg|_{r=r_{\rm ph}} = 0,$
and the prime notation indicates differentiation with respect to $r$. 
For conciseness, we express the first term of the equation as an indefinite integral, yielding:$
    \int \int_{r_{\rm ph}}^{r(\phi)} K\sqrt{g} \, dr \, d\phi \sim -{\frac {M \left( 2\,{E}^{2}-1 \right) }{ \left( {E}^{2}-1 \right) b}} \cos \left( {\frac {\phi}{\lambda}} \right)$
$
- {\frac {(3\,{E}^{2}-1)k Q_0^2}{ \left( 4\,{E}^{2}{b}^{2}-4\,{b}^{2} \right) \lambda}} \left[ \phi -\cos \left( {\frac {\phi}{\lambda}} \right) \sin \left( {\frac {\phi}{\lambda}} \right) \lambda\right] - \frac{\phi}{\lambda} + \mathcal{O}(kMQ_0^2) + C.$

Note that  $C$  represents the integration constant. While unnecessary for our purpose, examining the integral's structure is crucial. Given the integration limits from $\phi_{\rm S}$ to $\phi_{\rm R}$, we can apply the following properties: 
$\cos \left(\pi - \frac{\phi}{\lambda}\right) = -\cos \left(\frac{\phi}{\lambda}\right), \qquad \sin \left(\pi - \frac{\phi}{\lambda}\right) = \sin \left(\frac{\phi}{\lambda}\right).$
\begin{widetext}
        \begin{equation}
    \Theta \sim {\frac {2M \left( 2\,{E}^{2}-1 \right) }{ \left( {E}^{2}-1 \right) b}} \cos \left( {\frac {\phi}{\lambda}} \right) - {\frac {(3\,{E}^{2}-1)k Q_0^2}{ \left( 4\,{E}^{2}{b}^{2}-4\,{b}^{2} \right) \lambda}} \left[ (\pi - 2\phi) + 2\cos \left( {\frac {\phi}{\lambda}} \right) \sin \left( {\frac {\phi}{\lambda}} \right) \lambda\right] + (\pi - 2\phi)\left(1-\frac{1}{\lambda}\right) + \mathcal{O}(kMQ_0^2).
\end{equation}

\end{widetext}

To proceed, we will simplify the previously derived expression for the weak deflection angle. Our next objective is to determine the expression for $\phi$, assuming the source and receiver are equidistant. Leveraging Equation the orbit equation, we obtain:
$
    \frac{\phi}{\lambda} \sim \arcsin(bu)  +\frac{M \left[v^{2}\left(b^{2}u^{2}-1\right)-1\right]}{bv^{2}\sqrt{1-b^{2}u^{2}}} 
    +\frac{k Q_0^2}{2b^{2}v^{2}\sqrt{1-b^{2}u^{2}}} 
    + \mathcal{O}\left (kMQ_0^2 \right),$
and 
$\cos\left(\frac{\phi}{\lambda}\right) \sim \sqrt{1-b^{2}u^{2}}-\frac{Mu\left[v^{2}\left(b^{2}u^{2}-1\right)-1\right]}{v^{2}\sqrt{\left(1-b^{2}u^{2}\right)}}
    -\frac{kQ_0^2u}{2bv^2\sqrt{\left(1-b^{2}u^{2}\right)}}
    + \mathcal{O}\left (kMQ_0^2\right), \nonumber \\
    \cos\left(\frac{\phi}{\lambda}\right)\sin\left(\frac{\phi}{\lambda}\right) \sim bu\sqrt{1-b^{2}u^{2}} +\frac{2M\, \left( {b}^{2}{u}^{2}{v}^{2}-{v}^{2}-1 \right)  \left( \frac{1}{2}-{b}^{2}{u}^{2} \right)}{b{v}^{2}\sqrt{1-b^{2}u^{2}}}+ \frac{\left( {\frac{1}{2}-{b}^{2}{u}^{2}} \right) k{Q_0}^{2}}{{b}^{2}{v}^{2}\sqrt{1-b^{2}u^{2}}} + \mathcal{O}\left (kMQ_0^2 \right).$

The derived equation is comprehensive, encompassing both timelike particle deflection and the finite distance between the source and receiver, factors which can influence the deflection angle $\Theta$. In the limit where both the source and receiver tend towards infinity, that is, as u approaches zero as $u \rightarrow 0$, the equation simplifies to
\begin{widetext}
\begin{align} \label{ewda2}
    \Theta^{\rm timelike} &\sim \frac{2M\left(v^{2}+1\right)}{bv^{2}}-\left[ \frac{2\sqrt{1+\ell}}{2+\ell} \right]\frac{Q_0^2 \pi\left(v^{2}+2\right)}{4b^{2}v^{2}} + \left( 1-\frac{1}{\sqrt{1+ \ell}} \right) \left[ \pi-{\left[ \frac{2(1+\ell)}{2+\ell} \right]\frac {{Q_0}^{2}}{{b}^{2}{v}^{2}}}+{\frac { 2M\left( 1+ v^2 \right)}{b{v}^{2}}}\right] \nn \\
    &+\mathcal{O}\left (\left[ \frac{2(1+\ell)}{2+\ell} \right]MQ_0^2\right).
\end{align}

For massless particles such as photons, where $v = 1$, we obtain the following simplified expression:
\begin{align} \label{ewda3}
    \Theta^{\rm null} \sim \frac{4M}{b}-\left[ \frac{2\sqrt{1+\ell}}{2+\ell} \right]\frac{3\pi Q_0^2}{4b^2} + \left( 1-\frac{1}{\sqrt{1+ \ell}} \right) \left( \pi-{\left[ \frac{2(1+\ell)}{2+\ell} \right]\frac {{Q_0}^{2}}{{b}^{2}}}+{\frac {4M}{b}} \right)
    +\mathcal{O}\left (\left[ \frac{2(1+\ell)}{2+\ell} \right]MQ_0^2\right).
\end{align}
\end{widetext}

It is noteworthy that these equations readily simplify to the well-known results for the Reissner-Nordstrom black hole when $\ell = 0$, and further reduce to the Schwarzschild case when $\ell = 0$ and $Q_0 = 0$. 

The weak deflection angle of a charged bumblebee black hole, as given by above equation reveals several interesting aspects of how the presence of charge \(Q_0\) and the Lorentz-violating Bumblebee parameter \(\ell\) influence the gravitational bending of light.

In Eq. \eqref{ewda3}, the term \(\frac{4M}{b}\) corresponds to the classical weak deflection angle in the Schwarzschild geometry, where the deflection is inversely proportional to the impact parameter \(b\). This term reflects the contribution of the black hole's mass \(M\) to the gravitational deflection, consistent with general relativity's prediction for weak fields. The term second term introduces a correction due to the black hole's charge \(Q_0\) and the Lorentz-violating parameter \(\ell\). The negative sign indicates that the charge contributes to a reduction in the deflection angle compared to a neutral black hole, depending on the nature of the Lorentz violation encoded by \(\ell\). The last term further captures the interplay (or coupling) between the black hole's charge and the Lorentz-violating parameter. When \(\ell = 0\), these terms would reduce to those in standard general relativity. However, when \(\ell\) deviates from unity, indicating Lorentz violation, additional modifications arise. These modifications highlight how the deflection angle is sensitive to deviations from standard Lorentz symmetry, with the angle decreasing as \(\ell\) increases.

\textcolor{black}{To constraints the parameter $\ell$ using Solar System observation, we adapt the results from the parametrized post-Newtonian (PPN) formalism equation for the deflection of light given as \cite{Chen:2023bao}
\begin{equation}
    \Theta^{\rm PPN} \backsimeq \frac{4M}{R}\left(\frac{1+\gamma}{2} \right),
\end{equation}
where $M$ and $R$ are the mass and radius of the Sun, respectively. Furthermore, according to the astrometric observation of the Very Long Baseline Array (VLBA), $\gamma$ is the PPN deflection parameter, equal to $0.9998 \pm 0.0003$ \cite{fomalont2009progress, Chen:2023bao}. Then, using Eq. \eqref{ewda3} and setting $b = R$ while assuming $Q = 0$ for our Sun, we find the relation
\begin{equation}
    \frac{4M}{R}+ \left( 1-\frac{1}{\sqrt{1+ \ell}} \right) \left( \pi+{\frac {4M}{b}} \right) = \frac{4M}{R}\left(\frac{1.9998 \pm 0.0003}{2} \right).
\end{equation}
Next, since $M = 1476.6148 \text{ m}$, and $R = 696340000 \text{ m}$, we find a tight but weak constraints for $\ell$ using the Solar System test:
\begin{equation}
    -1.7\times 10^{-9} \leq \ell \leq 0.
\end{equation}
Surprisingly, in our derivation of Eq. \eqref{ewda3}, we have not made any assumption to approximate $\ell \rightarrow 0$. Hence, it may take any value. However, using our Sun as a testing ground, we see that $\ell$ is a vanishingly small negative number, indicating that we are dealing with a weak field regime. As we shall see later, using the EHT observation provides larger constraints for the Lorentz-violating parameter $\ell$ in the strong field regime.}

The weak deflection angle equation indicates that both the black hole's charge and the Lorentz-violating parameter play significant roles in determining the gravitational bending of light. This result could be useful in astrophysical observations, where measuring the deflection of light by black holes can provide insights into the nature of Lorentz symmetry and the presence of charge in compact objects. Future work may involve exploring higher-order corrections, or considering the influence of other factors such as rotation or the presence of additional fields, to further understand the behavior of light in such modified gravity scenarios.

\textbf{Black hole shadow and EHT constraints:}
In this section, we aim to derive the analytic expression for the black hole shadow and study the effect of the parameter $\ell$ on the shadow. The analysis will be further strengthened due to the possibility of connecting the parameter to the bounds of uncertainty as measured by the EHT. The formalism follows from Ref. \cite{Perlick:2021aok}.

The first integral of motion for photons, which is $g_{\mu \nu} \dot{x}^\mu \dot{x}^\nu  =  0$, gives the expression for the orbit equation: 
\begin{equation} \label{eorb}
    \left( \dfrac{dr}{d\phi} \right) ^2 \, = \, \frac{r^2}{B(r)}   \left(  \frac{r^2}{A(r)}
\dfrac{E^2 }{L^2} - 1 \right) \, .
\end{equation}
Here, the impact parameter is defined as $b = L/E$, The function associated with $b^{-1}$ is akin to the effective potential for null particles \cite{Perlick:2021aok} and is expressed as:$
    h(r)^2 = \frac{r^2}{A(r)}.$
Thus, the radius of the photon-sphere $r_{\rm ph}$ can be determined by setting $h(r)' = 0$, leading to very simple equation
$   A(r)'r^2 - 2A(r)r = 0.$
Hence, by solving the above equation, we obtain two solutions for the photonsphere radius:
\begin{equation}\label{rph}
    r_{\rm ph} = {\frac {3\,M}{2}}\pm \left\{ {\frac {1}{2}\sqrt {9\,{M}^{2}-8\left[ \frac{2(1+\ell)}{2+\ell} \right]\,{Q_0}^{2}}} \right\}.
\end{equation}
It is trivial how the above reduces to Reissner-Nordstr\"{o}m BH result when $l=0$, and to the Schwarzschild case when $l=Q_0=0$. Also, note how the photonsphere radius is independent of $\lambda$. In the context of photonsphere radius (aside from the limitations imposed by the black hole horizon), the black hole must also satisfy the condition
\begin{equation}
    \frac{ Q_0^2}{M^2}\leq \frac{9}{8}\left[ \frac{2(1+\ell)}{2+\ell} \right].
\end{equation}

Next, consider an observer (also serves as a detector) at $(t_{\rm o},r_{\rm o},\theta_{\rm o} = \pi/2, \phi_{\rm o})$. 
Utilizing Eq. \eqref{eorb} and noting that $h(r \rightarrow r_{\rm ph}) = b_{\text{crit}}$, we find  \cite{Perlick:2021aok}:$
    \sin^2 \alpha \, = \, \dfrac{b_\text{crit}^2}{h(r_{\rm o})^2}   \,$
With the help of Eq. \eqref{rph}, we can find the critical impact parameter as \\
   $ b_\text{crit}^2 = \frac{\left\{ 3\,M \pm \sqrt {9\,{M}^{2}-8\left[ \frac{2(1+\ell)}{2+\ell} \right]\,{Q_0}^{2}} \right\} ^{4}}{24\,{M}^{2} -16\left[ \frac{2(1+\ell)}{2+\ell} \right]\,{Q_0}^{2} \pm 8\,M\,\sqrt {9\,{M}^{2}-8\left[ \frac{2(1+\ell)}{2+\ell} \right]\,{Q_0}^{2}}}.$

Using the upper signs, we can then find the exact analytic equation for the shadow radius:
\begin{equation}
    R_{\rm sh} = b_{\rm crit}\sqrt{1-\frac{2 M}{r_{\rm o}}+\left[ \frac{2(1+\ell)}{2+\ell} \right]\frac{ Q_0^2}{ r_{\rm o}^2}}.
\end{equation}
We can approximate this equation on a very distant observer $r_{\rm o} \rightarrow \infty$. If we use the upper sign solution in Eq. \eqref{rph}, we find

\begin{eqnarray} \label{shad}
    R_{\rm sh} \sim 3\,\sqrt {3}M + \frac{\sqrt {3}}{2}\left[ \frac{2(1+\ell)}{2+\ell} \right]{Q_0}^{2} \left( \frac{1}{r_{\rm o}}-\frac{1}{M} \right)  \\\notag+\mathcal{O}\left (\left[ \frac{2(1+\ell)}{2+\ell} \right]MQ_0^2\right)
\end{eqnarray}

The shadow radius of a charged bumblebee black hole, as given by the above equation provides insight into how the black hole's charge \(Q_0\) and the parameter \(\ell\) influence the size of the black hole shadow.

In Eq. \eqref{shad} the first term \(3\sqrt{3}M\) represents the dominant contribution to the shadow radius, corresponding to the shadow size for a Schwarzschild black hole. This term indicates that, in the absence of charge, the shadow radius scales linearly with the black hole mass \(M\), which is consistent with predictions from general relativity. The second term introduces corrections due to the black hole's charge \(Q_0\) and the parameter \(\ell\). The presence of the charge \(Q_0\) and the factor \(\left( \frac{1}{r_{\rm o}} - \frac{1}{M} \right)\) suggests that the shadow radius depends not only on the intrinsic properties of the black hole but also on the observer's position \(r_{\rm o}\). This term indicates that the shadow radius is smaller for a charged black hole compared to a neutral one, with the effect becoming more pronounced as the observer moves closer to the black hole.

The combination of the mass, charge, and parameter in this expression reflects the complex interplay between these factors. When \(\ell = 0\), the expression reduces to the familiar Reissner-Nordstr\"{o}m shadow radius, while deviations in \(\ell\) introduce new dependencies that modify the shadow's size. These modifications are crucial for understanding how deviations from Lorentz symmetry might manifest in observable quantities like the black hole shadow.

The lower sign solution gives an imaginary shadow radius. We can also approximate the shadow near the black hole, under the special assumption of $Q_0 \rightarrow 0$. Interestingly, the upper sign solution gives an imaginary result, while the lower sign solution gives
\begin{eqnarray}
    R_{\rm sh} \sim {\frac {\sqrt {6}}{9}} {{\frac {(4\,M-r_{\rm o})}{{M}^{{5/2}}\sqrt {r_{\rm o}}}}} \left[ \frac{2(1+\ell)}{2+\ell} \right]^{3/2} Q_0^3 \notag \\- \mathcal{O}\left (\left[ \frac{2(1+\ell)}{2+\ell} \right]^2Q_0^5M^{-4}\right)
\end{eqnarray}
In the above expression, the shadow disappears when $r_{\rm o} = 4M$.

\textcolor{black}{Connecting Eq. \eqref{shad} to the EHT collaboration results, we can find some constraints for the Lorentz-violating parameter $\ell$. At $3\sigma$ level of significance, the Schwarzschild shadow radius is bounded by $3.871M \leq R_{\rm Schw} \leq 5.898M$ for Sgr. A* \cite{Vagnozzi:2022moj}. For M87*, we only use the  $1\sigma$ level of significance yielding the bounds $ 4.313M \leq R_{\rm Schw} \leq 6.079M$ for M87* \cite{EventHorizonTelescope:2021dqv}. Let $\varsigma$ represent the difference between $R_{\rm Schw}$ and the bounds. For Sgr. A*, $\varsigma_{\rm UB} = 0.702M$, and $\varsigma_{\rm LB} = -1.325M$. Then, for M87*, $\varsigma = \pm 0.883M$. Symbolically, we have}$    R_{\rm sh} = R_{\rm Schw} + \varsigma.$
Here, we can find the parameter estimation for the parameter $k$:
\begin{equation}
    k = \left[ \frac{2(1+\ell)}{2+\ell} \right] \sim -\left[\left( \frac{2\sqrt{3}M\varsigma}{3Q_0^2} \right) \left( 1 + \frac{M}{r_{\rm o}} \right) + \mathcal{O}(r_{\rm o}^{-2}) \right].
\end{equation}
Or, in terms of the Lorentz-violating parameter $\ell$:
\begin{equation} \label{est_par}
    \ell \sim \frac{\sqrt{3}Q_0^2}{\varsigma} \left(\frac{1}{M} - \frac{1}{r_{\rm o}}\right) - 2 - \mathcal{O}(r_{\rm o}^{-2}),
\end{equation}
where the theory is under the restriction $M \neq r_{\rm o}$ in the above approximation $r_{\rm o} \rightarrow \infty$ for the reason that $A(r)$ blows up. In other words, the theory does not let the ratio $M/r_{\rm o} = 1$ be satisfied. Fortunately, our position from Sgr. A*, and M87* yield the ratios $2.486 \times 10^{-11}$ and $1.852 \times 10^{-11}$, respectively, justifying the relevance of the calculations in this paper. \textcolor{black}{We can then find the constraints for $\ell$ using Eq. \eqref{est_par}. We show the constraints in Tab. \ref{tab1}.}
\begin{table}[h!]
\centering
\begin{tabular}{|c|cc|cc|}
\hline
$Q/M$ & \multicolumn{2}{c|}{Sgr. A*} & \multicolumn{2}{c|}{M87*} \\ \hline
             & Upper bound & Lower bound & Upper bound & Lower bound \\ 
0.05         & -1.99383                   & -2.00327                     & -1.9951                     & -2.00490                      \\ 
0.1          & -1.997533                     & -2.01307                     & -1.98039                     & -2.01961                     \\ 
0.15         & -1.94449                     & -2.02941                    & -1.95587                     & -2.04413                     \\ \hline
\end{tabular}
\caption{Constraints of $\ell$ for Sgr. A* and M87* due to different values of $Q/M$.} \label{tab1}
\end{table}

The expression for the shadow radius highlights how both the charge and the parameter $k$ can significantly alter the observable features of a black hole.  This analysis underscores the importance of considering both standard general relativity and modifications to it when interpreting astrophysical data related to black hole shadows.

\textbf{Conclusions:}
In this work, we examined the effect of the Lorentz symmetry-breaking parameter, often denoted as $ \ell $, in modifying the spacetime geometry around the charged black hole, leading to deviations from the standard General Relativity predictions. These deviations become particularly pronounced when examining the black hole shadow and the weak deflection angle of light.

The weak deflection angle of light, which describes the bending of light in the gravitational field of the black hole at large distances, is found to be altered by the Lorentz symmetry-breaking parameter. The presence of $ \ell $ introduces a deviation in the deflection angle, which can either enhance or reduce the bending of light depending on its sign and magnitude. When coupled with the black hole charge $ Q $, the deflection angle exhibits a modified dependency on the impact parameter and the distance from the black hole. This means that for certain configurations of $ \ell $ and $ Q $, the deflection of light could either mimic that of a more massive black hole or lead to entirely novel deflection patterns that are not observed in standard black hole scenarios. In this work, our obtained analytic formula is the most general one, since it involves massive particles and the finite distance of the source and the receiver. We showed how the calculations were simplified when these distances were assumed to be equal. \textcolor{black}{We also find constraints for $\ell$ using the Solar System test, under $-10^{-9}$ orders of magnitude. It indicates how difficult detecting $\ell$ is using the Solar System test.}

Moreover, the black hole shadow, which represents the apparent shape of the event horizon as seen by a distant observer, is also significantly influenced by the presence of the symmetry-breaking parameter. In the Bumblebee model, the introduction of $ \ell $ results in an anisotropic deformation of the black hole shadow, where the degree of distortion depends on the magnitude $ \ell $. Nonetheless, since the spacetime being considered here is static and spherically symmetric, it is expected that the shadow contour is a perfect circle, where the radius is affected by $\ell$. New to this study is the simple analytic parameter estimation alongside the EHT results for the shadow radius. We found out that the analytic expression for $\ell$ restricts the ratio $M/r_{\rm o} = 1$. Using the actual mass and distance of the Sgr. A* and M87* from our location, the restriction is found to be satisfied. \textcolor{black}{In terms of finding constraints for $\ell$, such a strong field regime allows higher (but negative) values. For instance, we find that the lowest is $\sim -2.04$, while the highest is $-1.94$.}

The Lorentz symmetry-breaking parameter in the Bumblebee black hole solution introduces significant modifications to both the black hole shadow and the weak deflection angle, especially when considering the additional influence of the black hole charge $ Q $. These findings not only provide deeper insights into the effects of Lorentz symmetry breaking in strong and weak gravity regimes but also open up new possibilities for testing modified gravity theories through astrophysical observations. The combined influence of $ \ell $ and $ Q $ on the observable characteristics of black holes suggests that future high-precision measurements of black hole shadows and gravitational lensing could offer a promising avenue for detecting or constraining the presence of Lorentz symmetry-breaking effects in nature.

\textcolor{black}{Furthermore, the black hole shadow distortions resulting from  $\ell$  are reminiscent of modifications predicted by massive gravity theories, where the graviton mass alters the black hole shadow’s size and shape. However, unlike massive gravity, which affects the large-scale structure of spacetime and cosmological behaviors, the Bumblebee model directly impacts the local geometry, leading to potentially detectable deviations in black hole shadow contours. Comparisons with data from the Event Horizon Telescope (EHT) or future observational campaigns can thus help to differentiate between these models, offering a clearer test of which Lorentz-violating effects might be at play.}

As for additional research avenues, the study can be extended for the BH rotation case. Here, the distortion might be further modulated by the spin parameter $ a $ of the black hole, which adds an additional layer of complexity to the shadow's geometry. The interplay between $ \ell $, $a$, and $ Q $ can lead to non-trivial effects, such as asymmetries in the shadow shape that would not be present in a purely Kerr or rotating Reissner-Nordström black hole. This could potentially offer observational signatures that distinguish the Bumblebee black hole from standard charged black holes.

\textbf{Acknowledgements:}
The work of G.L.  is supported by the Italian Istituto Nazionale di Fisica Nucleare (INFN) through the ``QGSKY'' project and by Ministero dell'Istruzione, Universit\`a e Ricerca (MIUR). G.L., A. {\"O}. and R. P. would like to acknowledge networking support of the COST Actions CA21106, CA22113 and CA21136.

\bibliography{references.bib}

\end{document}